\documentclass[aps,twocolumn,floats]{revtex4}
\usepackage{bm}
\usepackage{graphicx}
\usepackage{xcolor}

\begin{document}

\title{Hidden-symmetry-protected quantum pseudo-spin Hall effect in optical lattices }
\author{Jing-Min Hou$^1$}\email{jmhou@seu.edu.cn}
\author{Wei Chen$^2$}
\affiliation{$^1$Department of Physics,
Southeast University, Nanjing  211189, China}
\affiliation{$^2$College of Science, Nanjing University of Aeronautics and
Astronautics, Nanjing 210016, China}
\date{\today}
\begin{abstract}
We propose a scheme to realize a new $Z_2$ topological insulator in a square optical lattice. Different from the conventional topological insulator protected by the time-reversal symmetry,  here, the optical lattice possesses a novel hidden symmetry, which is responsible for the present $Z_2$ topological order.  With a properly defined pseudo-spin, such a topological insulator is characterized by the helical edge states which exhibits pseudo-spin-momentum locking,  so that it can be considered as a quantum pseudo-spin Hall insulator. The $Z_2$ topological invariant is derived and its experimental detection is discussed as well.

\end{abstract}
\maketitle

\section{Introduction}
 In the last decades, topological
 phases in condensed matters have attracted many attentions\cite{Hasan,Qi}.
 Conventionally, it was thought that matters were classified  according to symmetries based on Landau's theory\cite{Landau}.
  In 1980s, the discovery of integer quantum
Hall effect\cite{Klitzing}, which was firstly recognized to be
related to topology, changed the viewpoint of physicists on the
classification of matters\cite{Thouless}.   The discovery of the
time-reversal symmetry protected topological insulators stimulates
more interests on the topological
  matters protected by symmetries\cite{Kane1,
Kane2, Bernevig1,
Bernevig2,Konig,Fu,Moore,Roy,Fu2,Hsieh,Xia,Zhang,Chen,Hsieh2,Hsieh3}.
 Topological classification of matters are generally correlated with symmetries.
  Depending on the dimensionality and the
symmetry classes specified by time reversal symmetry and
particle-hole symmetry, gapped systems can be classified into ten
types of topological phases\cite{Schnyder,Kitaev}. Besides the
time-reversal and particle-hole symmetries, spatial symmetries, such
as point symmetry can protect a new kind of topological insulators,
called topological crystalline insulators\cite{Fu3,Hsieh4,CXLiu}. In
our previous work, we find a hidden symmetry in a square lattice,
which
 is response for the existence of the Dirac points\cite{Hou}. This kind of  hidden symmetry
is a discrete antiunitary symmetry with a  composite operator
consisting of translation, complex conjugation, and sublattice
exchange. A natural question is whether there exist topological
insulators protected by such kind of hidden symmetry. In this paper,
we give a positive answer for this question and develop a new kind
of topological insulators protected by such hidden symmetry.

  The
 development of optical lattice and cold atom techniques provide
versatile models, especially, some of which   are hardly realized in
solid real materials. Therefore,   cold atoms in optical lattices
become a platform to investigate and explore various kinds of
topological phases, especially those that are difficult to be
realized in the real materials. In recent years, many schemes have
been proposed to realize various topological phases with neutral
atoms in optical lattices. The the Harper
Hamiltonian and effective magnetic fields have been experimentally realized\cite{Miyake,Aidelsburger,Aidelsburger2,Struck1,Struck2}.
Resorting to staggered  effective magnetic fields, quantum anomalous Hall effect has been proposed to be
realized in honeycomb lattice\cite{Shao}, non-Abelian optical
lattices\cite{Goldman1}. For the case  with the time-reversal
symmetry, quantum spin Hall effect was proposed   to be realized
with neutral atoms  in
 optical   lattices\cite{Goldman2,Liu,Kennedy,Beri}.
The schemes have also been designed to realize
 the time-reversal symmetry protected three dimensional  topological
 insulators with neutral atoms in optical lattice\cite{Beri,Bermudez}.
As a result, the mature techniques of  cold atoms
and optical lattices  make it possible to realize the topological
insulators protected by the hidden symmetry.

 Here, we propose a scheme of cold atoms in an optical lattice that preserves
the hidden symmetry but breaks the time reversal symmetry due to the
existence of the hopping-accompanying phases. It is found that this
system supports the $Z_2$ topological insulators protected by the
hidden symmetry. In such an optical lattice,   a pseudo-spin operator can be defined, which has the
eigenvalues of $\pm 1$, corresponding  to the
pseudo-spin-up and pseudo-spin-down states. The pseudo-spin-up and
pseudo-spin-down states are related by the hidden symmetry operation. Therefore, we dub the two-dimensional topological insulators
as quantum pseudo-spin Hall (QPSH) insulator. The edge states are
helical, that is to say, the pseudo-spin-up and pseudo-spin-down states
 move along
the opposite directions along the edge of the lattice, which is just
the hallmark of the QPSH effect. We also define the hidden-symmetry
polarization, which is an integer modulo $2$, so that it is a $Z_2$ topological invariant. The
odd and even of the hidden-symmetry polarization correspond to the
QPSH insulator and the trivial band insulator, respectively. It should be pointed out that although the hidden symmetry plays a role in the QPSH
effect as the time reversal symmetry does in the QSH effect, the
hidden symmetry is distinctly from the time reversal symmetry. The
hidden symmetry includes a translation operator, so the hidden
symmetry operator is momentum-dependent in the concrete representation.

\begin{figure}[ht]
 \includegraphics[width=0.44\columnwidth]{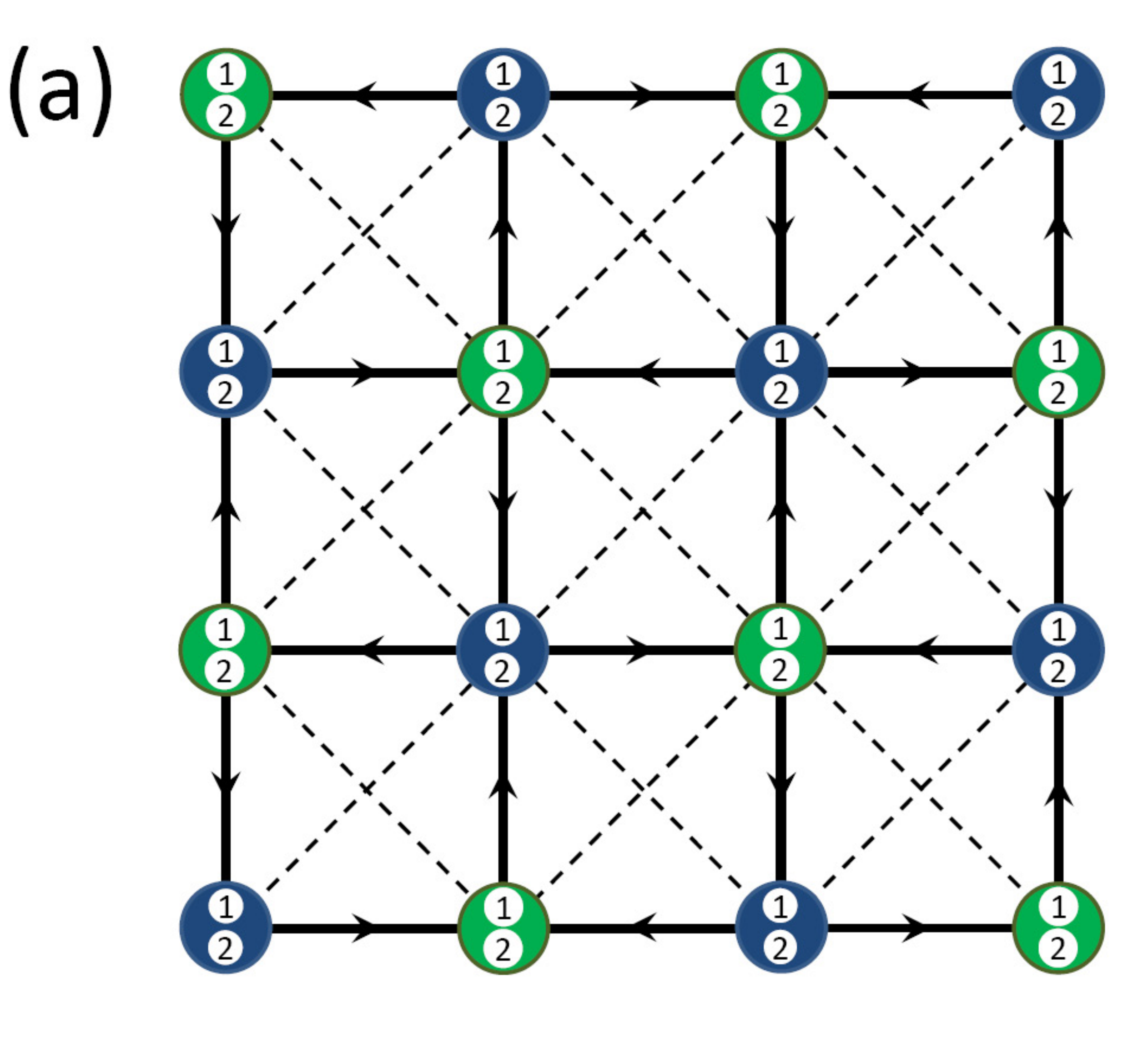}
 \includegraphics[width=0.44\columnwidth]{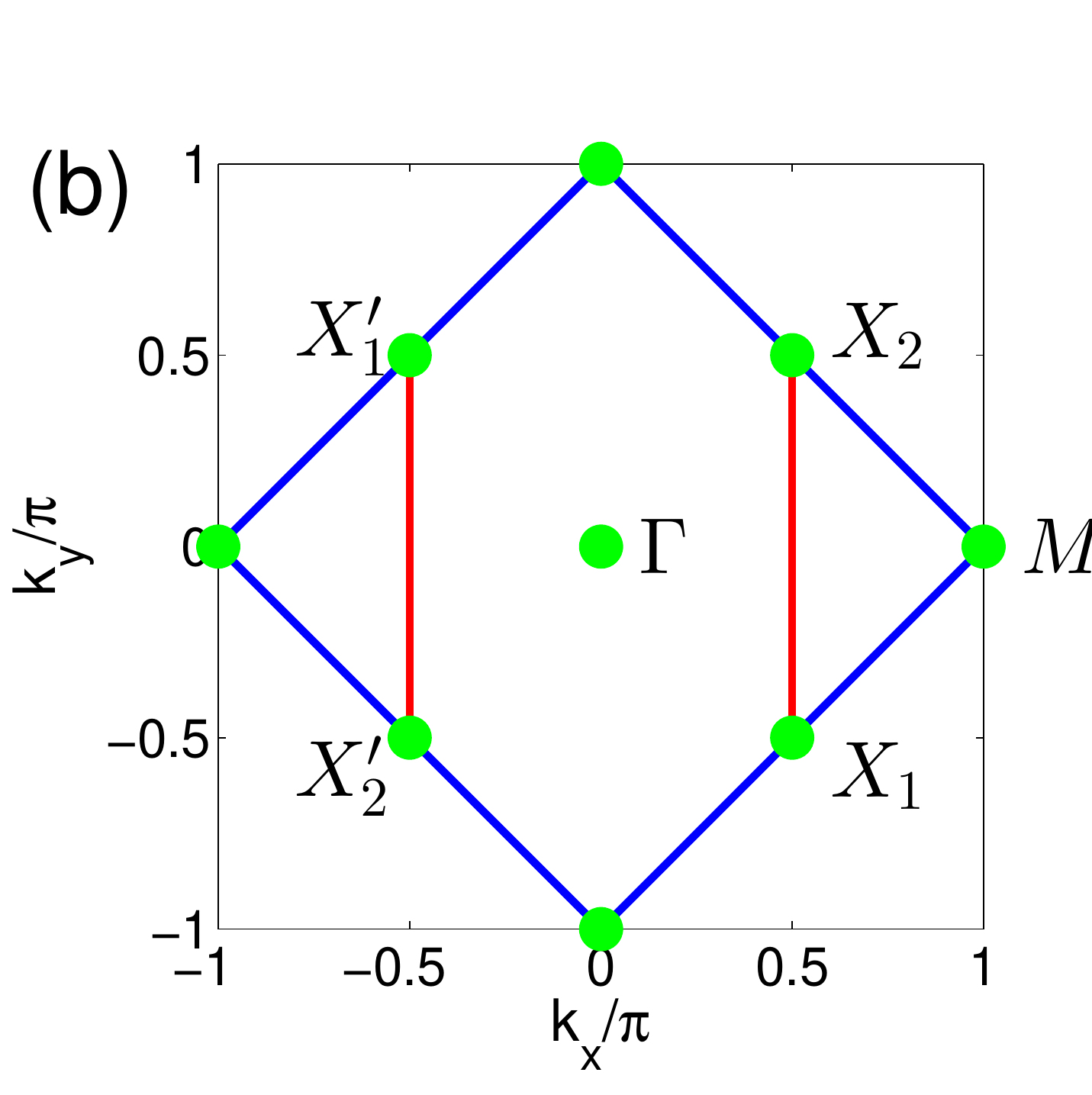}
\caption{(Color online).   (a) Schematic of the square optical
lattices and the designed phase factor (denoted by arrows). Here,
the solid and dashed lines represent the nearest and next-nearest
(diagonal) hopping, respectively; the green and blue circles represent
the lattice sites of sublattices $A$ and $B$, respectively; the numbers $1$ and $2$ in the circles denote  the intrinsic atomic states. (b) The
Brillouin zone, which is enclosed by the solid blue lines. Here, the
green   filled circles represent the high symmetry points. Along the
red lines, the $\Upsilon$-polarization is defined in the main text.
}\label{fig1}
\end{figure}

\section{Model}

Here, we consider a two-component (two-color) system on a square
lattice as shown in Fig.\ref{fig1}(a), where the arrows represent
the hopping-accompanying phases. Due to the presence of these
phases, the lattice is divided into  two sublattices $A$ and $B$.
 Each sublattice has the primitive
 lattice vectors $\mathbf{a}_1=(1,1)$ and
 $\mathbf{a}_2=(1,-1)$. In momentum
 space, the primitive reciprocal vectors are $\mathbf{b}_1=( \pi,\pi)$ and
$\mathbf{b}_2=(\pi, -\pi)$. The  corresponding Brillouin zone is
shown  in Fig.\ref{fig1}(b). The tight-binding Hamiltonian can be
written as $H=H_0+H_1+H_2$ with
\begin{eqnarray}
  H_{0}&=&-t\sum_{i\in A} [  e^{- i\gamma} a^\dag_{i}(-i\tau_y)
b_{i+ \hat{x} }  +  e^{- i\gamma} a^\dag_{i}(i\tau_y) b_{i- \hat{x}
}\nonumber\\
&&
 + e^{ i\gamma} a^\dag_{ i} (-i\tau_y) b_{i+\hat{y} } +  e^{
i\gamma } a^\dag_{ i} (i\tau_y) b_{i -\hat{y} } + {\rm H.c.}]
\end{eqnarray}
and
\begin{eqnarray}
  H_1&=&-t_1\sum_{i\in A} [ a^{\dag}_{i}\tau_z
a_{i+\hat{x}+\hat{y}}+a^{\dag}_{i} \tau_z
a_{i-\hat{x}-\hat{y}}\nonumber\\
&& +a^{\dag}_{i}\tau_z a_{i-\hat{x}+\hat{y}}+a^{\dag}_{ i}\tau_z a_{
i+\hat{x}-\hat{y}}] \nonumber\\
&& -t_1\sum_{i\in B} [ b^{\dag}_{i}\tau_z
b_{i+\hat{x}+\hat{y}}+b^{\dag}_{i} \tau_z
b_{i-\hat{x}-\hat{y}}\nonumber\\
&& +b^{\dag}_{i}\tau_z b_{i-\hat{x}+\hat{y}}+b^{\dag}_{ i}\tau_z b_{
i+\hat{x}-\hat{y}}]
\end{eqnarray}
and
\begin{eqnarray}
H_2=\lambda\sum_{i\in A}a^\dag_i\tau_z a_i+\lambda\sum_{i\in
B}b^\dag_i\tau_z b_i
\end{eqnarray}
where $a_i=[a_i^{(1)}, a_i^{(2)} ]^T$ and
$b_i=[b_i^{(1)},b_i^{(2)}]^T$ are the two-component  annihilation
operators destructing a particle at a lattice site of sublattice $A$
and $B$, respectively; $\tau_i (i=x,y,z)$   represent the Pauli
matrices in the color space; $t$ and $t_1$ represent the amplitudes
of hopping between the nearest lattice sites and between the
next-nearest lattice sites, respectively; $\lambda$ is an effective
magnetic field; $0<\gamma<\pi/2$ is a hopping-accompanying
phase. Here, the model   can be realized by applying $^6$Li or
$^{40}$K cold atoms trapped in an optical lattice. The color-switching hopping and the accompanying
phases of hopping   can be
realized with laser-assisted tunneling
techniques\cite{Aidelsburger2,Struck1,Struck2}.

\begin{figure}[ht]
\includegraphics[width=0.45\columnwidth]{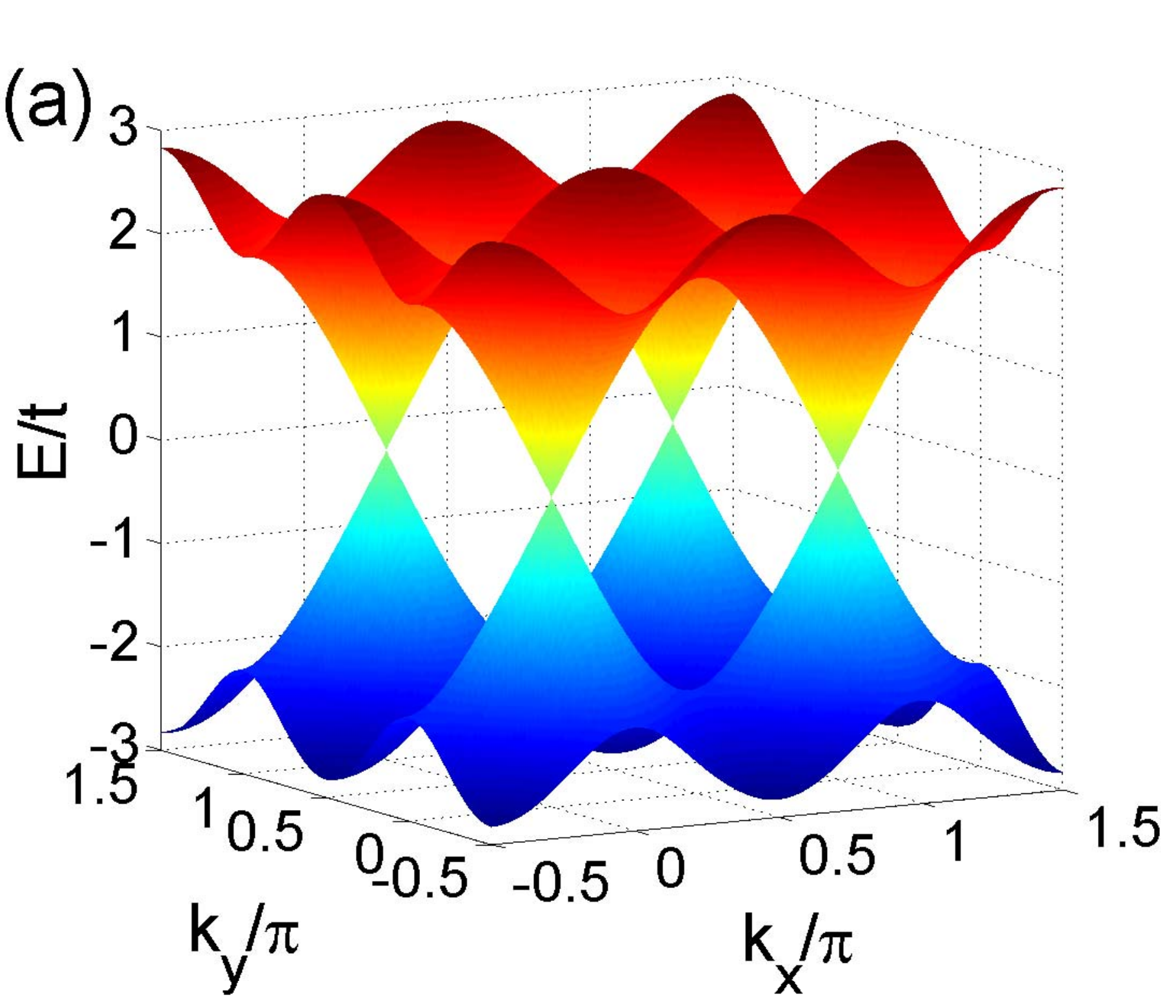}
\includegraphics[width=0.45\columnwidth]{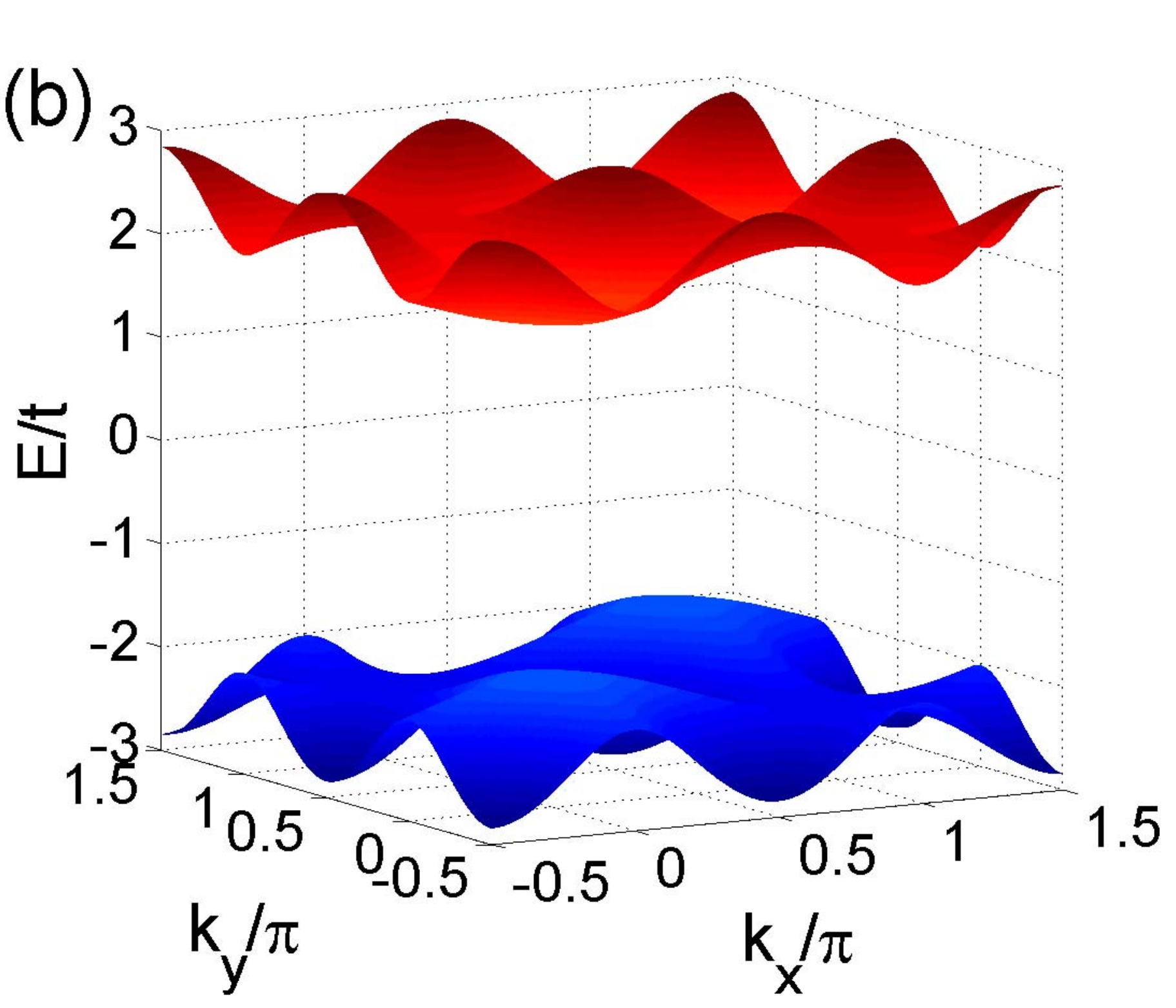}
\caption{(Color online).  The dispersion relation for (a)
$\gamma=\pi/4$, $t_1=0$ and $\lambda=0$; (b) $\gamma=\pi/4$,
$t_1=0.5t$ and $\lambda=0.3 t$. }\label{fig2}
\end{figure}

 Taking the Fourier's transformation on the
Hamiltonian $H$, we rewritten it as
$H=[a^{(1)\dag}_\mathbf{k},a^{(2)\dag}_\mathbf{k},
b^{(1)\dag}_\mathbf{k},b^{(2)\dag}_\mathbf{k} ]\mathcal{H}
(\mathbf{k})[a^{(1)}_\mathbf{k},a^{(2)}_\mathbf{k},b^{(1)}_\mathbf{k},b^{(2)}_\mathbf{k}]^T$
with
\begin{eqnarray}
{\cal H}( \mathbf{k}) =
 h_x(\mathbf{k})\sigma_x\otimes\tau_y +h_y(\mathbf{k})\sigma_y\otimes\tau_y+m(\mathbf{k})I\otimes
 \tau_z,\label{BH}
\end{eqnarray}
where $h_x(\mathbf{k})=-2t \cos\gamma(\sin k_x  +\sin k_y  )$ and
$h_y(\mathbf{k})=-2t \sin\gamma(\sin k_x -\sin k_y )$;
$m(\mathbf{k})=(\lambda-4t_1\cos k_x \cos k_y )$ is the mass term;
$\sigma_{x,y}$ represent the Pauli matrices in the sublattice space.
The dispersion relation is
$E(\mathbf{k})_\pm=\pm\sqrt{h_x(\mathbf{k})^2+h_y(\mathbf{k})^2+m(\mathbf{k})^2}$,
which is shown in Fig.\ref{fig2}. For this system, there are four
bands  and the valence and conduction bands are two-fold degenerate.
When the diagonal hopping terms $H_1$ and the effective magnetic
terms $H_2$ are absent, the mass term in the Bloch Hamiltonian
(\ref{BH}) vanishs. The corresponding  dispersion relation is shown
in Fig.\ref{fig2}(a), from which one finds that the Dirac points
occur at the points $\Gamma$ and $M$ in the Brillouin zone. When the
diagonal hopping terms $H_1$ and the effective magnetic terms $H_2$
are present, an gap opens as shown in fig.\ref{fig2}(b), then the
system turns into an insulator.

\section{Hidden Symmetry}
 It is easy to verify that the total
system $H$ preserves a hidden symmetry, i.e., $[H, \Upsilon]=0$ with the symmetry operator $\Upsilon$ defined as
\begin{eqnarray}
\Upsilon=(\sigma_x\otimes I )KT_{ {\hat{x}}}
\end{eqnarray}
where $T_{\hat{x}}$ is a translation operator which moves the
lattice by $\hat{x}$  along the $x$ direction; $K$ is the complex
conjugation operator; $\sigma_x$ is the Pauli matrix representing
the sublattice exchange and $I$ is the unit matrix in the color
space.

For the Bloch Hamiltonian, the corresponding transformation can be described by
\begin{eqnarray}
\Upsilon\mathcal{H}(\mathbf{k})\Upsilon^{-1}=\mathcal{H}(-\mathbf{k}).
\end{eqnarray}
Therefore, for each energy band, there always exists another energy band corresponding to the $\Upsilon$ transformed quantum states.
These two energy bands compose a pair of bands  related  to each other by the hidden symmetry $\Upsilon$, which are dubbed as the $\Upsilon$ pair bands.
The Bloch function is supposed to have the form
$\Psi_{\mathbf{k}}(\mathbf{r})=[u_{A,\mathbf{k}}^{(1)}(\mathbf{r}),
u_{A,\mathbf{k}}^{(2)}(\mathbf{r}),u_{B,\mathbf{k}}^{(1)}(\mathbf{r}),
u_{B,\mathbf{k}}^{(2)}(\mathbf{r})]^T
e^{i\mathbf{k}\cdot\mathbf{r}}$ in the coordinate representation. The
symmetry operator $\Upsilon$ acts on the Bloch function as follows
\begin{eqnarray}
\Upsilon\Psi_{\mathbf{k}}(\mathbf{r})&=&\left(\matrix{u^{(1)*}_{B,\mathbf{k}}(\mathbf{r}-\hat{x})e^{ik_x}\cr
u^{(2)*}_{B,\mathbf{k}}(\mathbf{r}-\hat{x})e^{ik_x}\cr
u^{(1)*}_{A,\mathbf{k}}(\mathbf{r}-\hat{x})e^{ik_x}\cr
u^{(2)*}_{A,\mathbf{k}}(\mathbf{r}-\hat{x})e^{ik_x}}\right)e^{-i\mathbf{k}\cdot\mathbf{r}}
=\Psi'_{\mathbf{k}'}(\mathbf{r}).\label{Bloch2}
\end{eqnarray}
 Because $\Upsilon$  is the symmetry operator of the system,
$\Psi'_{\mathbf{k}'}(\mathbf{r})$ must be a Bloch wave function of
the system. Thus, we obtain $\mathbf{k}'=-\mathbf{k}$,
$u^{(i)}_{A,\mathbf{k}'}(\mathbf{r})=u^{(i)*}_{B,\mathbf{k}}(\mathbf{r}-\hat{x})e^{ik_x}$
and
$u^{(i)}_{B,\mathbf{k}'}(\mathbf{r})=u^{(i)*}_{A,\mathbf{k}}(\mathbf{r}-\hat{x})e^{ik_x}$
with $i=1,2$. From Eq.(\ref{Bloch2}), it is easy to show that the
operator $\Upsilon$ has the effect when acting on  wave vectors
 as  $ \Upsilon: \mathbf{k}\rightarrow -\mathbf{k}=\mathbf{k}'$.
If $\mathbf{k}'=\mathbf{k}+\mathbf{G} $, where $\mathbf{G}$ is
the reciprocal lattice vector, then we can say that $\mathbf{k}$ is
a $\Upsilon$-invariant point in momentum space. We find four
distinct $\Upsilon$-invariant points in the Brillouin zone as
\begin{eqnarray}
\Gamma=(0,0), M=( \pi, 0), X_{1,2}=( \pi/2, \pm \pi/2).
\end{eqnarray}
For the hidden symmetry operator $\Upsilon$, we have
$\Upsilon^2=T_{2 {\hat{x}}}$, which has the representation based on
the Bloch wave functions as $\Upsilon^2=e^{-i2\mathbf{k}\cdot
{\hat{x}}} $. Thus,  we have $\Upsilon^2=-1$ at the
$\Upsilon$-invariant  points $X_{1,2}$ while $\Upsilon^2=1$ at
$\Gamma$ and $M$. Since $\Upsilon$ is an antiunitary operator, it is
straightforward to show that $\Upsilon$ protected degeneracy must occur
at the points $X_{1,2}$\cite{Hou}.

\section{Quantum pseudo-spin Hall effect}
We define an operator as
\begin{eqnarray}
S=\sigma_z\otimes \tau_z
\end{eqnarray}
which has two eigenvalues $\xi=\pm 1$. Thus, the operator $S$ can be considered as a pseudo-spin operator.  It is easy to verify that the pseudo-spin operator $S$ commutes with the Bloch Hamiltonian (\ref{BH}),
i.e., $[\mathcal{H}(\mathbf{k}),S]=0$, so the Bloch Hamiltonian and the pseudo-spin operator have common eigenstates. That is to say,
  every Bloch wave function can have a fixed eigenvalue of $S$ with $+1$ or $-1$.

When the gap opens, there exist two kinds of insulators, which are topologically distinct. When the system turns from one kind of insulator into the other, a topological phase transition happens.
We can manifest this scenario with band inversions.

When the mass term $m(\mathbf{k})$ is absent, the energy bands has
two distinct Dirac points at $\Gamma $ and
$M$ in the Brillouin zone. Around the Dirac points,
the Bloch Hamiltonian can be linearized  as the form
\begin{eqnarray}
\mathcal{H}_{\Gamma,M}(\mathbf{k})=\sum_{ij}v^{\Gamma,M}_{ij}p_i\sigma_j\otimes\tau_y
\end{eqnarray}
with
\begin{eqnarray}
v^{\Gamma}=\left(\matrix{-2t\cos\gamma&-2t\sin\gamma\cr
-2t\cos\gamma&2t\sin\gamma}\right)
\end{eqnarray}
and
\begin{eqnarray}
v^{M}=\left(\matrix{2t\cos\gamma&2t\sin\gamma\cr
-2t\cos\gamma&2t\sin\gamma}\right)
\end{eqnarray}
where $\mathbf{p}$ is the relative wave vector from the Dirac points. About the Dirac points, we find  $w=\textrm{sgn}(\det [v^{\Gamma,M}])=\pm 1$,
which has opposite  values for the two distinct Dirac points at $\Gamma$ and $M$.

\begin{figure}[ht]
\includegraphics[width=0.8\columnwidth]{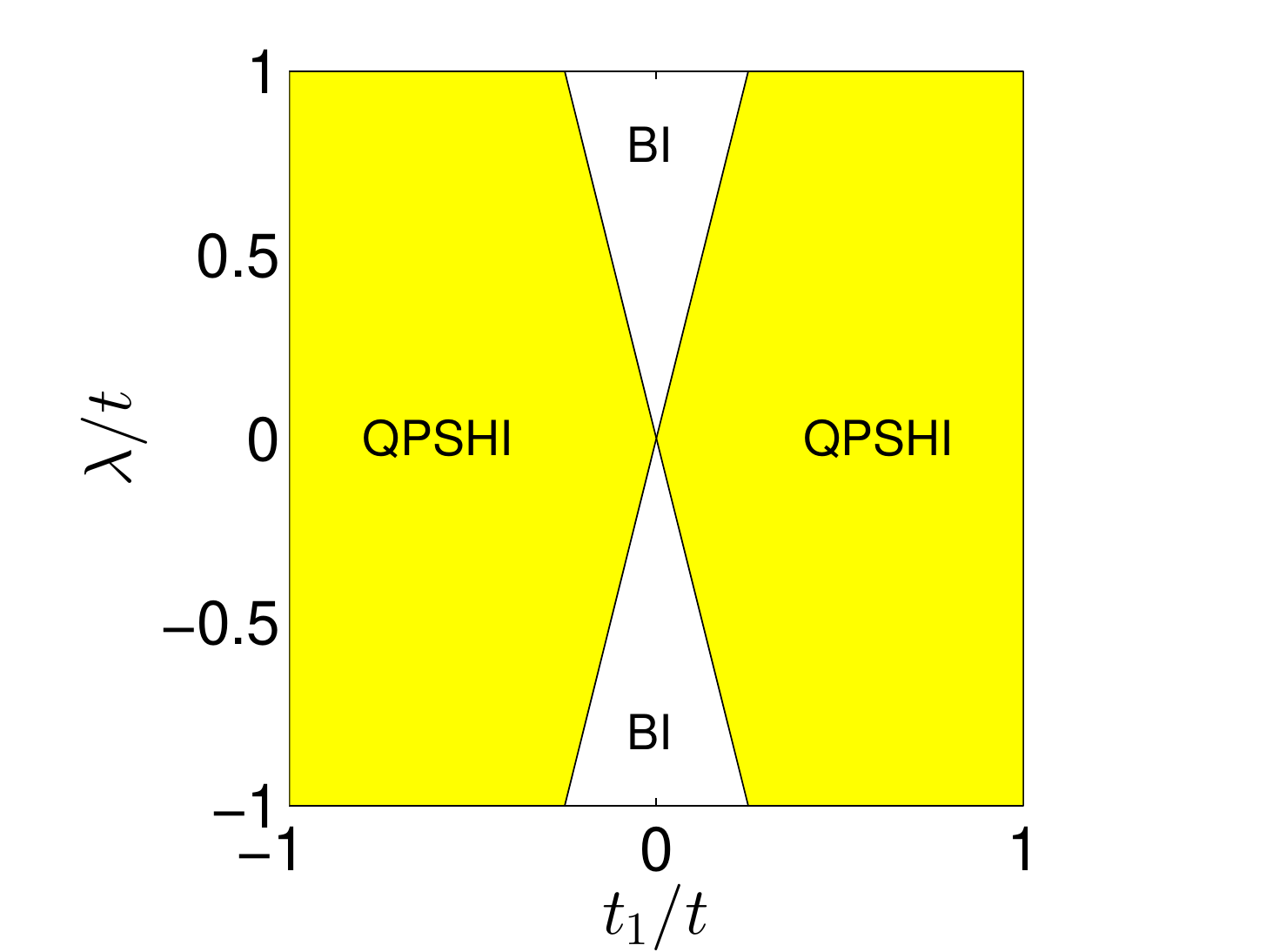}

\caption{(Color online).   The phase diagram. Here, QPSHI and BI
denote   quantum pseudo-spin Hall insulators and conventional band
insulators, respectively.  }\label{fig3}
\end{figure}

As we mentioned above, when the diagonal hopping terms $H_1$ and the effective magnetic terms $H_2$ are present, a gap opens and the sysmtem turns into an insulator for half-filling case.
We can find that the mass term in the Bloch Hamiltonian have different sign  in different parameter ranges. There are four cases as follow: (i) when
$\lambda-4t_1>0$ and $\lambda+4t_1>0$, the mass terms at two Dirac
points are both positive; (ii) when
$\lambda-4t_1<0$ and $\lambda+4t_1<0$, the mass terms at two Dirac
points are both negative; (iii)   when  $\lambda-4t_1<0$ and
$\lambda+4t_1>0$,  the mass  at  $\Gamma$
point is negative and the one at $M$ point still
remains positive; (iv) when  $\lambda-4t_1>0$ and
$\lambda+4t_1<0$ are satisfied,  the mass at  $\Gamma$
point is positive and the one at $M$ point is negative. For the two former cases, the mass terms at the points $\Gamma$ and $M$ have the same sign, the system is a conventional band insulator.
 In contrast, for the two latter cases,  they have opposite signs, the
system is a  topological insulator. The conventional band insulators and topological insulators can be distinguished by the edge states.
When the mass term at one of the points change sign, that is to say, a band inversion happens, which implies that  a topological  phase transition occurs. Based on the signs of the mass terms at the
Dirac points $\Gamma$ and $M$, we obtain the phase diagram as shown in Fig.\ref{fig3}.

\begin{figure}[ht]
\includegraphics[width=0.45\columnwidth]{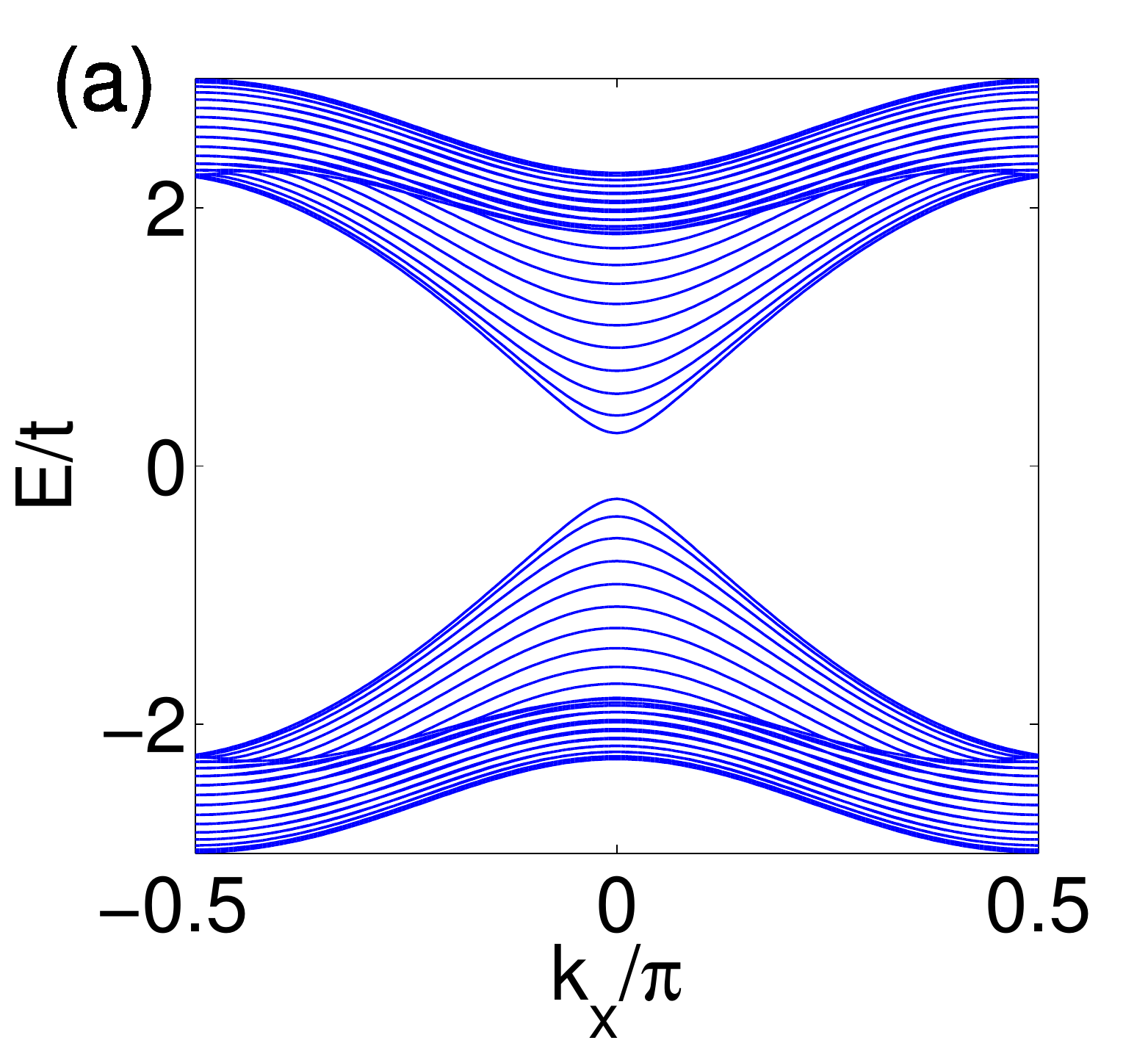}
\includegraphics[width=0.45\columnwidth]{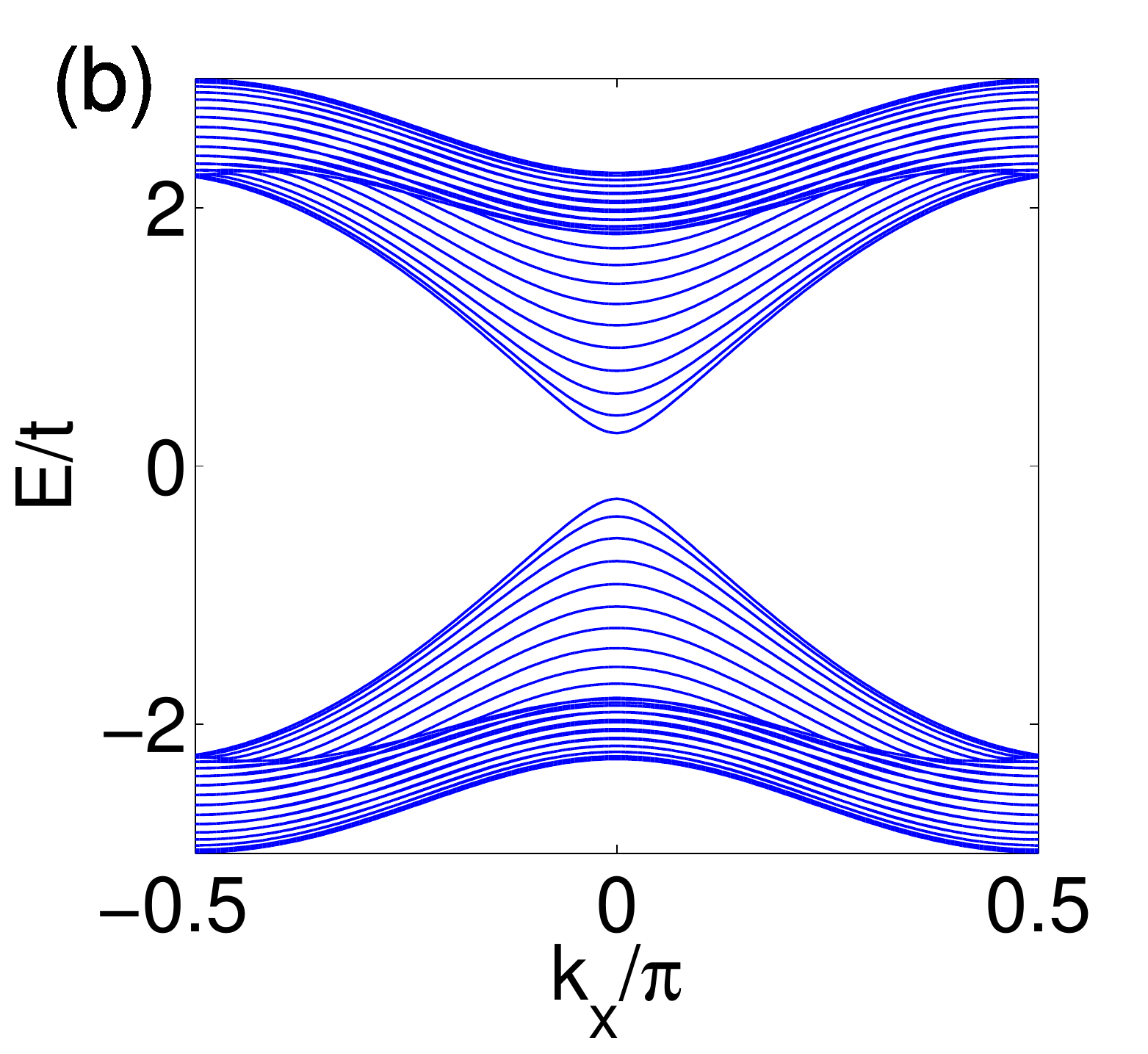}
\includegraphics[width=0.45\columnwidth]{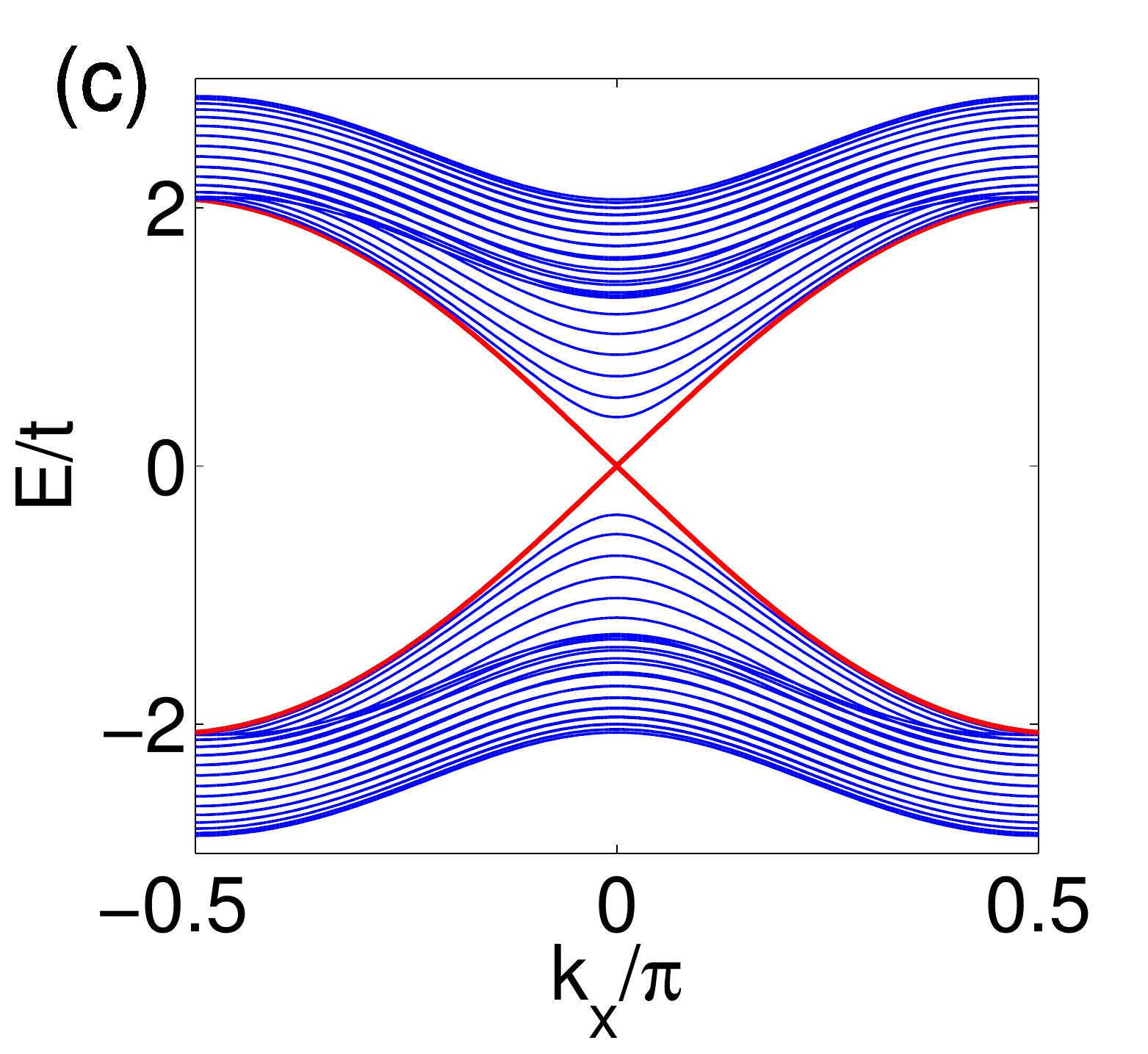}
\includegraphics[width=0.45\columnwidth]{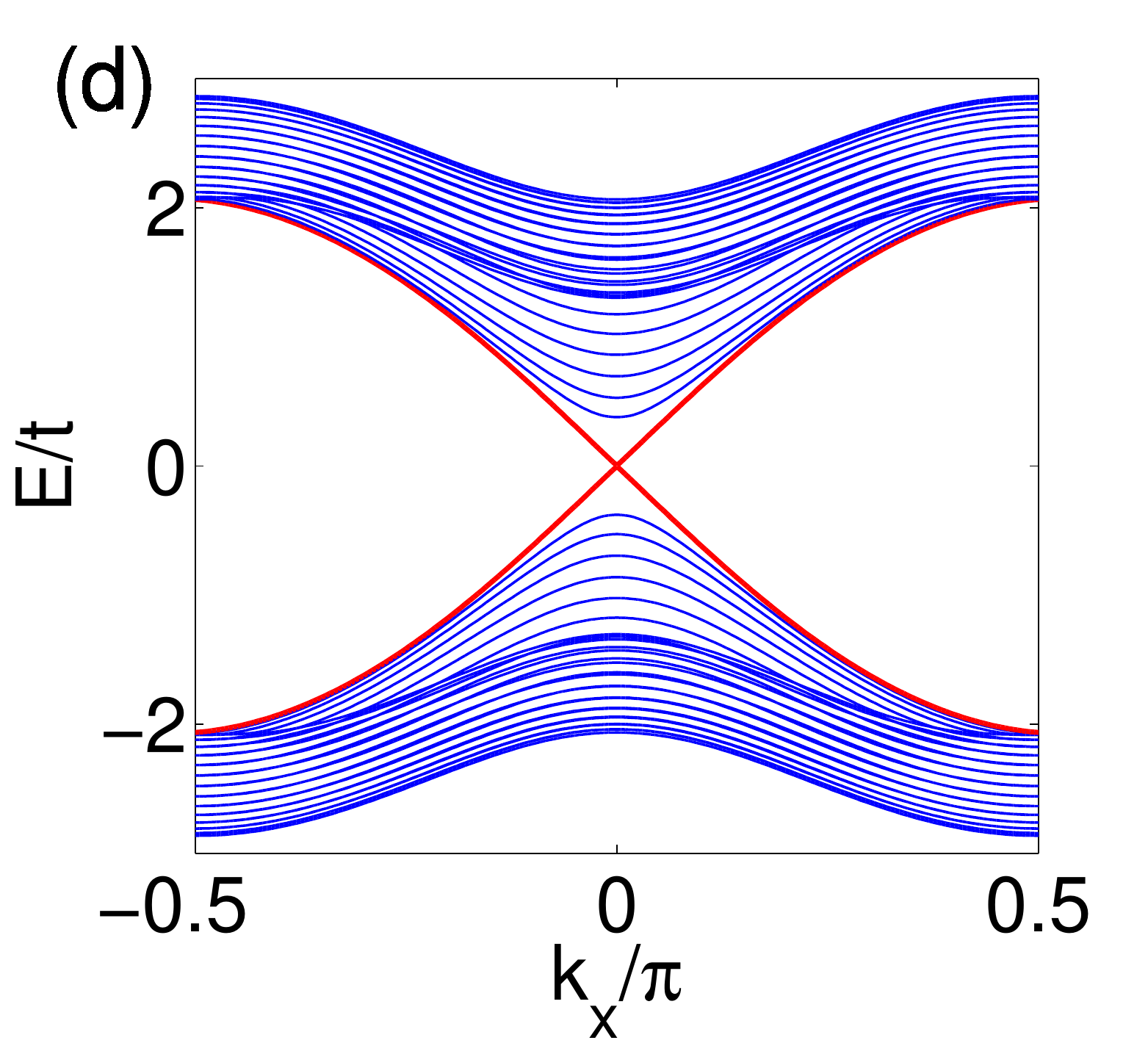}
\caption{(Color online).  The dispersion relations of a strip geometry
(a) for (i) case with $t_1=0.2t$ and $\lambda=t$, (b)   for (ii) case with $t_1=-0.2t$ and $\lambda=-t$, (c) for (iii) case with $t_1=0.2t$ and $\lambda=0.5 t$
(d) for (iv) case with $t_1=-0.2t$ and $\lambda=-0.5 t$.
}\label{edge}
\end{figure}

We obtain the edge states by diagonalizing the  Hamiltonian of
 a  strip  geometry. The dispersion relations of a strip geometry are shown in Fig.\ref{edge}. For (i) and (ii) parameter ranges,
  there are no  edge states  as shown in Figs.\ref{edge}(a) and (b), which confirms   that (i) and (ii) parameters correspond to the conventional band insulators.
  For (iii) and (iv) parameter ranges, there is a pair  of helical  edge states at each surface
 transversing the band gap as shown in Figs.\ref{edge}(c) and (d), which indicates the existence of topological insulators.

 We find that the edges states  $|\psi_{\pm,k_x}\rangle$ of the strip geometry  are eigenstates of the operator $S=\sigma_z\otimes
 \tau_z$ with the eigenvalue $\xi=\pm 1$, i.e. $S|\psi_{\pm,k_x}\rangle=\pm |\psi_{\pm,k_x}\rangle$. It is easy to verify that the operator $S$ anticommutes with the hidden symmetry operator $\Upsilon$, i.e., $[S,\Upsilon]_+=0$, so that we have
 \begin{eqnarray}
 \Upsilon|\psi_{\xi,k_x}\rangle=e^{i\alpha}|\psi_{-\xi,-k_x}\rangle
 \end{eqnarray}
 where $\alpha$ is a phase depending on the gauge. That  is to say,
  the particles in different edge
 states on the same boundary move in the opposite direction with different eigenvales of $S$. We
 can consider the operator $S$ as the pseudo-spin operator. Thus,
 the  edge states are pseudo-spin-momentum locking. The fact that the edge states
 are related by the hidden symmetry operator $\Upsilon$  is a solid evidence that the edge states are protected by the hidden symmetry $\Upsilon$.

\section{  $Z_2$ topological invariant}
 It is well known that the
time-reversal symmetry can lead to the existence of the $Z_2$
two-dimensional topological insulators, i.e., the quantum spin Hall
insulators. Here, the system considered here preserves a hidden
symmetry, which is also a discrete  antiunitary symmetry. Similar to
the time-reversal symmetry leading to  the quantum spin Hall
insulators, the hidden symmetry can also lead to a new kind of $Z_2$
topological insulators.

Similar to the definition of time-reversal polarization for the
time-reversal symmetry protected spin quantum Hall insulator\cite{Fu06,Fu07}, we can
define the $\Upsilon$ polarization for the insulators preserving the
$\Upsilon$ symmetry.

 The Bloch wave functions occupied band can be written as
 $|\Psi_{n,\mathbf{k}}\rangle=e^{i\mathbf{k}\cdot\mathbf{r}}|u_{n,\mathbf{k}}\rangle$,
 where $|u_{n,\mathbf{k}}\rangle$ is the the cell-periodic eigenstate
 of the Bloch Hamiltonian $\mathcal{H}(\mathbf{k})$.
 The Berry connection matrix
 \begin{eqnarray}
 \mathbf{a}_{mn}=-i\langle
 u_{m,\mathbf{k}}|\nabla_\mathbf{k}|u_{n,\mathbf{k}}\rangle
 \end{eqnarray}
 For the hidden symmetry $\Upsilon$, we define a matrix as
\begin{eqnarray}
w_{mn}(\mathbf{k})=\langle
u_{m,-\mathbf{k}}|\Upsilon|u_{n,\mathbf{k}}\rangle\label{wmatrix}
\end{eqnarray}
which is antisymmetric at the $\Upsilon$-invariant degenerate points $X_{1,2}$.

For the present model with half filling, there are two occupied bands, which compose the $\Upsilon$ pair bands. For the occupied $\Upsilon$ pair bands,
we can define the charge polarization in terms of the Berry connection along a direction in the Brillouin zone (see red lines in Fig.\ref{fig1} (b)) as
\begin{eqnarray}
P_\rho=\frac{1}{2\pi}\left[\int_{X_1}^{X_2}\mathbf{A}(\mathbf{k})\cdot
d\mathbf{k}+\int_{X_1'}^{X_2'}\mathbf{A}(\mathbf{k})\cdot
d\mathbf{k}\right]
\end{eqnarray}
where $\mathbf{A}(\mathbf{k})$ is defined as
$\mathrm{tr}[\mathbf{a}(\mathbf{k})]$. For each occupied band,   the partial
charge polarization is defined as
\begin{eqnarray}
P_i=\frac{1}{2\pi}\left[\int_{X_1}^{X_2}\mathbf{a}_{ii}(\mathbf{k})\cdot
d\mathbf{k}+\int_{X_1'}^{X_2'}\mathbf{a}_{ii}(\mathbf{k})\cdot
d\mathbf{k}\right]
\end{eqnarray}
For the $\Upsilon$ pair bands,
we can also define the $\Upsilon$ polarization as
\begin{eqnarray}
P_\Upsilon&=&P_1-P_2=2P_1-P_\rho\nonumber\\
&=&\frac{1}{2\pi}\int_{X_1}^{X_2}[\mathbf{A}(\mathbf{k})-\mathbf{A}(-\mathbf{k})]\cdot
d\mathbf{k}-\frac{i}{\pi}\log\frac{w_{12}(X_2)}{w_{12}(X_1)}\nonumber\\
&=&\frac{1}{i\pi}\log\left[\frac{\sqrt{w_{12}(X_1)^2}}{w_{12}(X_1)}\cdot\frac{w_{12}(X_2)}{\sqrt{w_{12}(X_2)^2}}\right]
\label{UpsilonPor}
\end{eqnarray}
which is an interger and   only defined modulo
$2$ due to the  ambiguity of the log. The argument of the log has only two values $\pm 1$ associated with the even and odd values of $P_\Upsilon$, respectively.
Therefore, we can rewrite Eq.(\ref{UpsilonPor}) as
\begin{eqnarray}
(-1)^{P_\Upsilon}
=\prod_{i=1}^2\frac{\textrm{Pf}[w(X_i)]}{\sqrt{\textrm{det}[w(X_i)]}}
\label{Z2}
\end{eqnarray}
The $Z_2$ topological invariant can be defined as $P_\Upsilon$ modulo $2$. When $P_\Upsilon$ is odd or even, the system is a topological insulator or a trivial band insulator.
Eq.(\ref{Z2}) gives a disticnt defination of the $Z_2$ topological invariant. However, evaluating the $Z_2$ topological invariant with Eq.(\ref{Z2}) requires a continuous gauge from the point  $X_1$ to the $X_2$ in the Brillouin zone, which is a difficult task for numerical
 calculations. Fortunately, the $Z_2$ topological invariant can be obtainedy by calculating the Berry gauge
 potential and
 the Berry curvature proposed by   Fukui and  Hatsugai\cite{Fukui}  or by evaluating the non-Abelian Berry connection derived from the Wannier function center proposed by Yu, et al.\cite{Yu}.

\section{Techniques for experimental detection of the $Z_2$ topological invariant}
Conventionally, the topological invariants in solid materials are
detected by measuring the transport properties. However, this method
is infeasible for cold atomic systems. Thus, other methods to identify
topological insulators have been proposed and even performed
experimentally. One method is to detect edge states by Bragg spectroscopy\cite{Goldman3} or direct imaging\cite{Goldman4}. An important  progress in measuring of topological invariant is the experiment realization of directly measuring the Zak phase by using a combination of Bloch
oscillations with Ramsey interferometry\cite{Atala}. Base on  the same experimental techniques, a scheme has been designed to measure the Chern number in two-dimensional lattices\cite{Abanin}. Later, the scheme was generalized to detect   $Z_2$ topological
invariants based the same experimental techniques\cite{Grusdt}. Therefore, these methods can also be applied to detect the topological order in the quantum pseudo-spin Hall insulator in our study.

\section{ Conclusion} In summary, we have proposed a scheme  to realize a new kind of  $Z_2$ topological insulator in a square optical lattice. Such nontrivial topological phase is protected by  the hidden symmetry $\Upsilon$ that is a composite   antiunitary symmetry
consisting of translation, complex conjugation, and sublattice
exchange. The hidden symmetry is intrinsically different time-reversal symmetry. Because of the inclusion of translation operation, the hidden symmetry is momentum-dependent in  a specific representation and there are only two symmetry-invariant points in the two-dimensional Brillouin zone instead of four symmetry-invariant points as the case of  time-reversal symmetry.
Based on the hidden symmetry, the pseudo-spin was defined and the $Z_2$ topological invariant was derived, for which the new topological insulator was dubbed as quantum pseudo-spin Hall  insulator.  Through numerical calculations, helical edge states was found for the non-trivial topological phase. Helical edge states are pseudo-spin-momentum locking and two edge states moving in opposite directions form the $\Upsilon$ pairs, which is a solid evidence of the new $Z_2$ topological insulator is  protected by  the hidden symmetry $\Upsilon$.  The detection of  the $Z_2$ topological invariant through the techniques of cold atoms and optical lattices was also discussed. Our work opens a perspective to search for new classes of topological phases by studying the hidden  symmetries of the lattices.

\begin{acknowledgments}
 J.M.H. acknowledges the support from the National Natural Science Foundation of China under Grant
No. 11274061; W.C. acknowledges the supports from the National Natural Science Foundation of China
under Grant No. 11504171, the Natural Science Foundation of Jiangsu Province, China under Grants
No. BK20150734, and the Project funded by China Postdoctoral Science Foundation under Grants No.
2014M560419 and No. 2015T80544.
\end{acknowledgments}

\end{document}